\documentclass[mathleft
]{an}
\usepackage{graphicx}
\usepackage{times}
\usepackage{natbib}
\bibpunct{(}{)}{;}{a}{}{,}
\newcommand{\kms}{km~s$^{-1}$}
\begin{document}

\Pagespan{789}{}
\Yearpublication{2010}%
\Yearsubmission{2010}%
\Month{11}%
\Volume{999}%
\Issue{88}%

\title{The Bar and Spiral Structure Legacy (BeSSeL) Survey:\\Mapping the Milky Way with VLBI Astrometry}
\author{Andreas Brunthaler\inst{1}\fnmsep\thanks{Corresponding author: \email{brunthal@mpifr-bonn.mpg.de}\newline} 
\and Mark J. Reid\inst{2}
\and Karl M. Menten\inst{1}
\and Xing-Wu Zheng\inst{3}
\and Anna Bartkiewicz\inst{4}
\and Yoon K. Choi\inst{1}
\and Tom Dame\inst{2}
\and Kazuya Hachisuka\inst{5}
\and Katharina Immer\inst{1,2} 
\and George Moellenbrock\inst{6}
\and Luca Moscadelli\inst{7}
\and Kazi L.J. Rygl\inst{1,8}
\and Alberto Sanna\inst{1} 
\and Mayumi Sato\inst{9} 	
\and Yuanwei Wu\inst{10}
\and Ye Xu\inst{10} 
\and Bo Zhang\inst{1}
}
\titlerunning{Mapping the Milky Way}
\authorrunning{Brunthaler, Reid, Menten et al.}
\institute{Max-Planck-Institut f\"ur Radioastronomie, Auf dem H\"ugel 69, 53121 Bonn, Germany
\and Harvard-Smithsonian Center for Astrophysics, 60 Garden Street, Cambridge, MA 02138, USA
\and Department of Astronomy, Nanjing University, Nanjing 210093, China
\and Torun Centre for Astronomy, Nicolaus Copernicus University, Gagarina 11, 87-100 Torun, Poland
\and Shanghai Astronomical Observatory, 80 Nandan Road, Shanghai, China
\and National Radio Astronomy Observatory, Socorro, NM, USA
\and INAF, Osservatorio Astrofisico di Arcetri, Largo E. Fermi 5, 50125 Firenze, Italy
\and Istituto di Fisica dello Spazio Interplanetario (IFSI-INAF), Via del Fosso del Cavaliere 100, 00133 Roma, Italy
\and Department of Astronomy, Graduate School of Science, The University of Tokyo, Tokyo 113 0033, Japan
\and Purple Mountain Observatory, Chinese Academy of Sciences, Nanjing 210093, China
}
\received{28 Feb 2011}
\accepted{}
\publonline{later}


\abstract{%
Astrometric Very Long Baseline Interferometry (VLBI) observations of maser 
sources in the Milky Way are used to map the spiral structure of our Galaxy 
and to determine fundamental parameters such as the rotation velocity
($\Theta_0$) and curve and the distance to the Galactic center (R$_0$). Here, 
we present an update on our first results, implementing a recent change in the 
knowledge about the Solar motion. It seems unavoidable 
that the IAU recommended values for R$_0$ and $\Theta_0$ need a substantial 
revision. In particular the combination of 8.5 kpc and 220 \kms\, can be ruled
out with high confidence. Combining the maser data with the distance 
to the Galactic center from stellar orbits and the proper motion of Sgr\,A*
gives best values of R$_0$ = 8.3 $\pm$ 0.23 kpc and $\Theta_0$ = 239 or 246 
$\pm$ 7 \kms, for Solar motions of V$_ \odot$ = 12.23 and 5.25 \kms, 
respectively. Finally, we give an outlook to future observations in the Bar and
Spiral Structure Legacy (BeSSeL) Survey.}

\maketitle

\section{Introduction}
The Milky Way is a barred spiral galaxy, as seen from observations of CO and 
H{\sc i} gas \citep[e.g.][]{Burton1988, DameHartmannThaddeus2001}, and
star counts \citep[e.g.][]{BenjaminChurchwellBabler2005}. However, our location
in the Galaxy makes it difficult to determine the number and positions of 
spiral arms, the length and position of the central bar, and the rotation 
curve. As a result, even the most fundamental parameters of the Milky Way, 
such as the distance to the Galactic center R$_0$ and the rotation speed 
$\Theta_0$ are still not known with high accuracy. However, these values
are not only important for Galactic astronomy, but also for a wide range
of different fields. This includes the interpretation of the proper motions
of the Magellanic Clouds \citep{ShattowLoeb2009,BeslaKallivayalilHernquist2010,DiazBekki2011,RuzickaTheisPalous2010,Peebles2010} and galaxies in the Andromeda 
subgroup \citep{BrunthalerReidFalcke2005,BrunthalerReidFalcke2007,vanderMarelGuhathakurta2008}, the motion of the Sun relative to the cosmic microwave 
background \citep[e.g.][]{LoebNarayan2008}, and even on the interpretation of 
dark matter direct detection experiments \citep{McCabe2010,Foot2010}.

In recent years, many large scale surveys have covered the Galactic Plane in 
all wave bands from radio to gamma rays. However, almost all of these surveys 
are all only two dimensional, and using them to construct a three dimensional 
model of the 
Milky Way is not trivial, mainly due to large uncertainties in distance 
measurements \citep{GeorgelinGeorgelin1976,HouHanShi2009}. These also affect the
interpretation
of these surveys since most astrophysical quantities, such as
linear size, mass, and luminosities, strongly depend on the distance to the
object.

\section{Galactic distances}

In Galactic astronomy, the {\it kinematic distance} is very commonly used. Here,
a distance is deduced from a measured line-of-sight velocity, a rotation 
model of Milky Way, and the assumption that the object has no peculiar motion. 
Apart from a distance ambiguity in the inner Galaxy (near and far kinematic
distance), errors in the rotation model and peculiar motions can lead to very 
large errors in the kinematic distance. One extreme example is the star forming
region G9.62+0.20. It is located at a distance 5.2 $\pm$ 0.6 kpc 
\citep{SannaReidMoscadelli2009}, while the near and far kinematic distances 
place the source at 0.5 and 16 kpc, respectively. This source shows a very 
large peculiar motion which may be induced by the central bar and is 
responsible for the large error in the kinematic distance.

Certainly, the trigonometric parallax is the most fundamental method, since it
is based on pure geometry without any astrophysical assumptions except the 
well known orbit of the Earth about the Sun. However, the measurement of an 
accurate trigonometric parallax requires  extremely high astrometric precision.
Friedrich Wilhelm Bessel was able to measure the first stellar parallax of the 
star 61 Cygni \citep{Bessel1838MNRAS, Bessel1838AN}. A huge step forward came
with the ESA's Hipparcos satellite \citep{PerrymanLindgrenKovalevsky1997}. It 
provided astrometric accuracies of the order of 1 milliarcsecond, which allows
distance estimates in the Solar neighborhood out to 100 pc with 10\% accuracy,
i.e. a very small portion of the Milky Way. Following the footsteps of 
Hipparcos, ESA's new Gaia mission \citep{PerrymandeBoerGilmore2001} will be 
launched in late 2012 to provide astrometry of up to 1 billion stars in the 
Milky Way with parallax accuracies up to 7 $\mu$as if it achieves 
specifications, a factor 100 better than Hipparcos. Although Gaia will 
revolutionize our view of the Milky Way, the 
strong extinction by gas and dust in the Galactic plane and in particular the 
spiral arms will prevent Gaia from making significant progress on the spiral
structure of the Milky Way. Since radio waves are not affected by interstellar
extinction, radio astronomy can come to the rescue and fill the gaps.

\section{VLBI Astrometry}
Recently developed calibration techniques for Very Long Baseline Interferometry
(VLBI) have improved the accuracy of astrometric VLBI observations 
significantly. When these techniques are applied to VLBI networks like the NRAO
Very Long Baseline Array (VLBA) in the US, the European VLBI Network (EVN) in
Europe and China, or the VLBI Exploration of Radio Astrometry (VERA) array in 
Japan, it is now possible to measure parallaxes of radio sources with 
accuracies of the order of 10~$\mu$as, comparable to the expected accuracy of 
Gaia. 


Possible target objects for VLBI astrometry are either radio continuum sources
or strong maser emission. Most radio stars are relatively weak, and the 
sensitivity of current instruments limits the detection to sources within a 
few hundred parsec \citep[e.g.][]{LoinardTorresMioduszewski2007,TorresLoinardMioduszewski2007, MentenReidForbrich2007}. However, sensitivity upgrades
of existing VLBI instruments \citep[e.g.][]{UlvestadRomneyBrisken2010} and 
eventually the Square Kilometer Array \citep{FomalontReid2004} will extend
the accessible range to even weaker and more distant sources.

Molecules with maser lines suitable for astrometry 
are hydroxyl (OH), methanol (CH$_3$OH), water (H$_2$O), and silicon monoxide 
(SiO). Since OH masers at a low frequency of 1.6 GHz are strongly affected by
interstellar scattering and ionospheric effects, they are not the optimal
targets for VLBI astrometry.  SiO masers at 43 GHz are less affected by 
scattering and the ionosphere, but these observations are more vulnerable to 
tropospheric phase errors and require excellent weather conditions. 
Nevertheless, both masers have been used to measure accurate distances to 
evolved stars \citep[e.g.][]{vanLangeveldeVlemmingsDiamond2000, ChoiHirotaHonma2008}. 

The most suitable targets are methanol (6.7 and 12.2 GHz) and water (22 GHz) 
masers. They are very strong and are found in high mass star forming regions 
(HMSFRs) mainly in the spiral arms of the Galaxy. The catalog of 6.7 GHz 
methanol masers by \cite{PestalozziMinierBooth2005} lists more than 500 sources,
and many more were found recently with Arecibo \citep{PandianGoldsmithDeshpande2007}, Effelsberg \citep{XuLiHachisuka2008}, and in particular in the methanol 
multi beam survey on the Parkes telescope \citep{CaswellFullerGreen2010,GreenCaswellFuller2010}. Water masers can be even stronger and \cite{ValdettaroPallaBrand2001} list more than 1000 sources in the Galaxy.

\begin{table}
\centering
{\begin{tabular}{lcl}
Source & Parallax& Reference\\
& [$\mu$as]&\\
\hline
W3(OH)          & 512 $\pm$ 10 & \cite{XuReidZheng2006}\\
Orion Nebula    & 2425 $\pm$ 35 & \cite{MentenReidForbrich2007}\\ 
S\,269          & 189 $\pm$ 16 & \cite{HonmaBushimataChoi2007}\\
VY\,CMa         & 876 $\pm$ 76 & \cite{ChoiHirotaHonma2008}\\
NGC\,281        & 355 $\pm$ 30 & \cite{SatoHonmaKobayashi2008}\\
S\,252          & 476 $\pm$ 6 & \cite{ReidMentenBrunthaler2009}\\ 
G232.6+1.0      & 596 $\pm$ 35 & \cite{ReidMentenBrunthaler2009}\\ 
Cep\,A          & 1430 $\pm$ 80 & \cite{MoscadelliReidMenten2009}\\   
NGC\,7538       & 378 $\pm$ 17 & \cite{MoscadelliReidMenten2009}\\ 
W51\,IRS2       & 195 $\pm$ 71 & \cite{XuReidMenten2009}\\
G59.7+0.1       & 463 $\pm$ 20 & \cite{XuReidMenten2009}\\
G35.2-0.7       & 456 $\pm$ 45 & \cite{ZhangZhengReid2009}\\
G35.2-1.7       & 306 $\pm$ 45 & \cite{ZhangZhengReid2009}\\
G23.0-0.4       & 218 $\pm$ 17 & \cite{BrunthalerReidMenten2009}\\  
G23.4-0.2       & 170 $\pm$ 32 & \cite{BrunthalerReidMenten2009}\\ 
G23.6-0.1       & 313 $\pm$ 39 & \cite{BartkiewiczBrunthalerSzymczak2008}\\
WB\,89-437      & 167 $\pm$ 6 & \cite{HachisukaBrunthalerMenten2009}\\ 
IRAS00420+5530  & 470 $\pm$ 20 & \cite{MoellenbrockClaussenGoss2009}\\ 
\end{tabular}
\caption{List of 18 sources which were used to fit a new model of the Galaxy in \cite{ReidMentenZheng2009}.}
\label{tab:sources}}
\end{table}

Parallaxes using the 6.7 GHz methanol transition have been reported so far only 
with the EVN \citep{RyglBrunthalerReid2010,RyglBrunthalerMenten2010}. while the
VLBA can currently observe only the 12.2 GHz methanol and 22 GHz water maser 
lines. One example is the distance to W3(OH), a high mass star forming region 
in the Perseus spiral arm of the Milky Way. Two independent parallax 
measurements with the VLBA of bright methanol \citep{XuReidZheng2006} and water 
\citep{HachisukaBrunthalerMenten2006} masers yielded consistent distance 
estimates of 1.95 $\pm$ 0.04 and 2.04 $\pm$ 0.07 kpc.

Another example that demonstrates the high quality of VLBI parallax 
measurements is the Orion Nebula. Four independent measurements with two 
different instruments give, within their joint errors, consistent results. 
\cite{SandstromPeekBower2007} observed one radio star with the VLBA and obtained
a distance of 389$^{+24}_{-21}$ pc, while \cite{HirotaBushimataChoi2007} used 
VERA to measure the parallax of water masers in the nebula (437 $\pm$ 19 pc).
The most accurate measurements of 414 $\pm$ 7 pc \cite[][VLBA, 4 radio stars]{MentenReidForbrich2007} and 418 $\pm$ 6 pc \cite[][VERA, SiO maser]{KimHirotaHonma2008} also agree with each other and with the modeling of the orbit of the 
binary $\Theta^1$\,Ori\,C from near-infrared interferometry
\citep[][434$\pm$12 pc]{KrausBalegaBerger2007}.

\section{A new model for the Milky Way}

\subsection{Measurements so far}
\begin{center}
\begin{figure}
{\includegraphics[width=83mm]{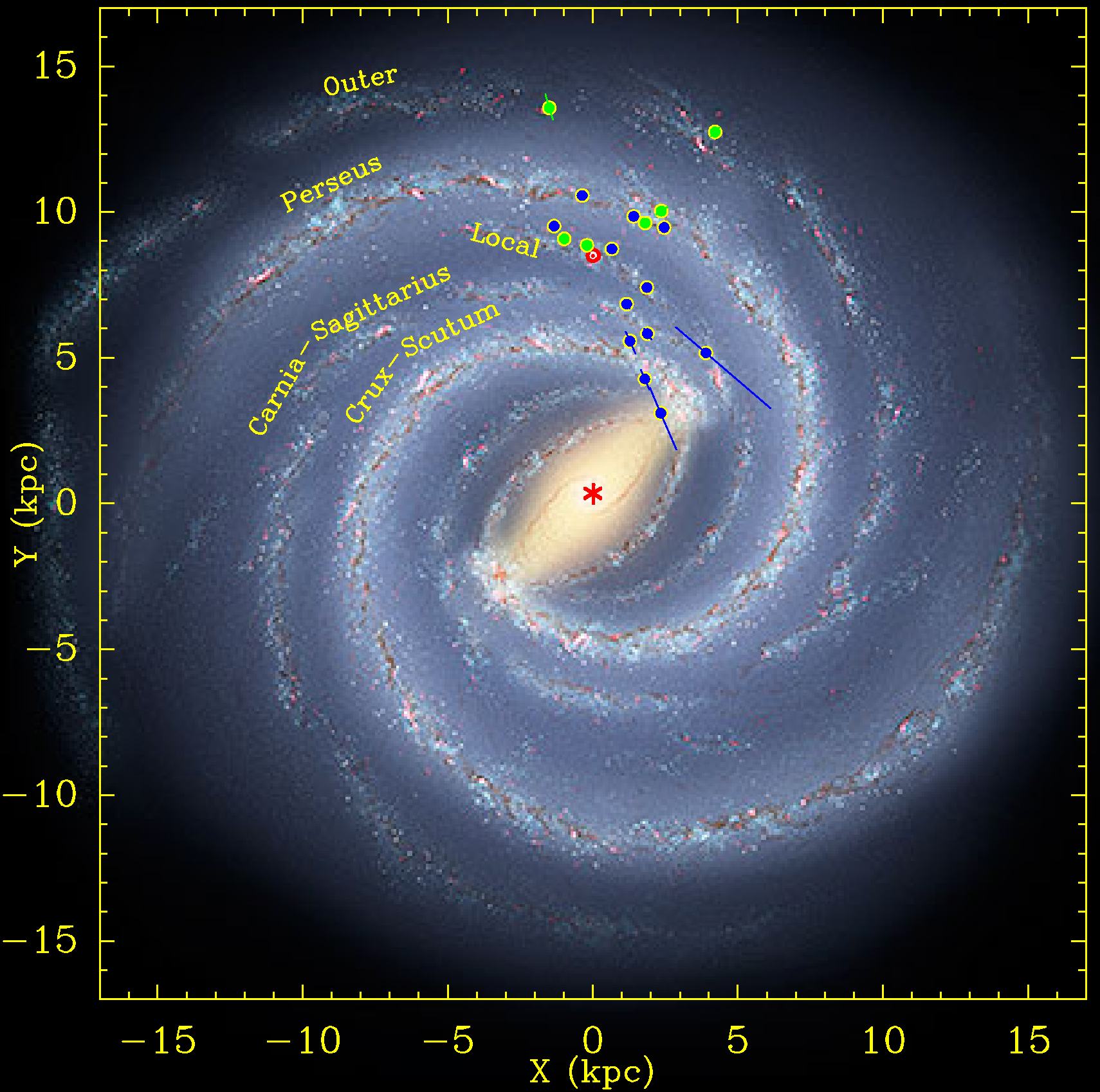}
\caption{Artist conception of the Milky Way (R. Hurt: NASA/JPL-Caltech/SSC) with the positions of the 18 sources from Table~\ref{tab:sources}. Sources in blue are 12.2 GHz methanol masers measured with the VLBA, while the green circles indicate H$_2$O masers observed with the VLBA or VERA and the Orion Nebula. The red symbols mark the position of the Galactic center and the Sun. Error bars are shown for all sources, but smaller
than the symbol size for most sources \citep[taken from][]{ReidMentenZheng2009}.}
\label{fig:sources}}
\end{figure}
\end{center}

In \cite{ReidMentenZheng2009}, we used 18 sources, with published parallaxes 
at that time (see Table~\ref{tab:sources} and Fig.~\ref{fig:sources}), to 
analyze the spiral structure and the rotation of the Milky Way. We were already 
able to locate the Local arm and the Perseus spiral arm from Galactic 
longitudes $l$ = 122$^\circ$  -- 190$^\circ$. The measured pitch angle of the 
Perseus spiral arm of $16^{\circ} \pm 3^{\circ}$ favors four rather than two 
spiral arms for the Galaxy. Furthermore, we find that most of the HMSFRs are 
closer than their kinematic distance. Note that the same data set has been also
analyzed by \cite{BovyHoggRix2009}, \cite{McMillanBinney2010}, and 
\cite{BobylevBajkova2010}.

\subsection{Solar Motion}
Astrometric observations not only yield the distance but also the proper motion
of a source. Together with the known position and line-of-sight velocity, 
this gives the full 6 dimensional phase space information for the observed 
sources. However, all measurements are heliocentric. Therefore, the peculiar 
motion of the Sun relative to the LSR is needed to convert the measured
heliocentric into Galactocentric motions.
For over a decade, the values of U$_\odot$= 10.00 $\pm$ 0.36, V$_\odot$= 5.25 
$\pm$ 0.62, and  W$_\odot$= 7.17 $\pm$ 0.38 \kms~for the peculiar motion of the
Sun derived from Hipparcos data by \cite{DehnenBinney1998} have been widely 
used. When using this Solar motion, we find that the HMSFRs rotate on average 
15 \kms\, slower than the Milky Way \citep{ReidMentenZheng2009}. 

This controversial result has sparked new interest in the Solar motion and 
\cite{Binney2010} argues that this slower rotation is partly induced by 
an erroneous value of V$_\odot$. \cite{Mignard2000} and 
\cite{PiskunovKharachenkovRoesler2006} already favored a significantly higher 
value of V$_\odot$, in the range of $\sim$12 \kms. This higher value, which is 
similar to the pre-Hipparcos value \cite[e.g.][]{MihalasBinney1981}, has been 
also found in a number of recent studies \citep{FrancisAnderson2009,SchoenrichBinneyDehnen2010,CoskunogluAkBilir2011}, while the values of U$_\odot$ and 
W$_\odot$ have not changed considerably \cite[see also][]{FuchsDettbarnRix2009}.
Although the issue of the V$_\odot$ component of the Solar motion is probably 
still not solved, it seems likely that the value of V$_\odot$ is probably 
closer to 12 \kms\,than to 5 \kms.

\subsection{Revised values}

The analysis in \cite{ReidMentenZheng2009} was based on the old Hipparcos
value of the Solar motion from \cite{DehnenBinney1998}. Here we give an update 
using the new values of \cite{SchoenrichBinneyDehnen2010}, which are U$_\odot$ 
= 11.10 $\pm$ 1.2, V$_\odot$= 12.24 $\pm$ 2.1, and W$_\odot$= 7.25 $\pm$ 0.6 
\kms. Using this value, the HMSFRs rotate $\sim$ 8 $\pm$ 2 \kms\, slower than 
the Milky Way. By fitting our measurements to a model of the Galaxy, we can 
also estimate the distance to the Galactic center R$_0$ and the circular
rotation speed $\Theta_0$. Assuming a flat rotation curve, we get  values of
R$_0$ = 8.4 $\pm$ 0.6 kpc and $\Theta_0$ = 247 $\pm$ 16 \kms. A similar
analysis allowing for different rotation curves yields results in the range of 
R$_0$ = 7.9 -- 9 kpc and $\Theta_0$ = 223 -- 280 \kms. However, the ratio of
these two values is much better constrained with a value of $\Theta_0$/R$_0$ = 
29.4 $\pm$ 0.9 \kms\, kpc$^{-1}$. Clearly, the number of sources is currently 
not large enough to constrain different rotation curves.

The slower average rotation of the HMSFRs together with the higher overall
rotation of the Milky Way also explains why most of the kinematic distances
are larger than the true distances. A description that takes into account
these two effects and gives (in general) more accurate kinematic distances,
and a more realistic distance uncertainty estimate is also presented in 
\cite{ReidMentenZheng2009}. It should be noted, however, that these revised 
kinematic distances can be still unreliable, because sources may have very 
large peculiar motions.

\subsection{Independent measurements of R$_0$ and $\Theta_0$/R$_0$}
The distance to the Galactic center R$_0$ has been the target of numerous 
investigations over the last decades \cite[see][for a review]{Reid1993}.
Most measurements in the range between 7.5 and 8.5 kpc and direct geometric 
distance estimates are rare. The accurate determination of stellar orbits in 
the Galactic center yields now consistent values from two groups of 8.4 $\pm$ 
0.4 kpc \citep{GhezSalimWeinberg2008} and 8.33 $\pm$ 0.35 kpc 
\citep{GillessenEisenhauerTrippe2009}. A trigonometric parallax to water masers
in Sgr\,B2, a star forming region located within 150 pc from the Galactic
center, also yields a consistent value of R$_0$ = 7.9$^{+0.8}_{-0.7}$ kpc, but
with a larger uncertainty \citep{ReidMentenZheng2009b}. Further observations 
of the water masers in Sgr\,B2 will improve the accuracy to a value comparable
or even better than from the stellar orbits. 

The ratio of $\Theta_0$ and R$_0$ is known with better than 1\% accuracy from 
the proper motion of Sgr\,A* \citep{ReidBrunthaler2004}. This motion in the 
Galactic plane of 6.379 $\pm$ 0.026 mas~yr$^{-1}$ is a combination of Galactic
rotation and Solar motion, and corresponds to a value for 
($\Theta_0$+V$_\odot$)/R$_0$ of 30.24 \kms\,kpc$^{-1}$. Using the Solar motion 
from \cite{SchoenrichBinneyDehnen2010}, and assuming the 
distance of 8.4 kpc, yields a value of $\Theta_0$/R$_0$ = 28.79 $\pm$ 0.26 
\kms\,kpc$^{-1}$, where the uncertainty is dominated by the uncertainty in the 
Solar motion. This value depends only weakly on the assumed R$_0$ (in the 
correction of the Solar motion), and does not change by more than 0.7\%, even 
for extreme values of R$_0$ of 7.5 or 9.5 kpc. This value is perfectly
consistent with our results from the independent maser parallaxes but
significantly larger ($>$11 $\sigma$) than the IAU value of 25.88 from the 
combination of R$_0$ = 8.5 kpc and $\Theta_0$ = 220 \kms. Hence, one requires 
an R$_0$ of less than  7.7 kpc to have a rotation velocity of 220 \kms. 

\subsection{Conclusion and best values for R$_0$ and $\Theta_0$}
The measured maser parallaxes, and the combination of the stellar orbits with 
the proper motion of Sgr\,A* provide independent and consistent evidence for
a higher rotation velocity of the Milky Way. Therefore, it seems unavoidable 
that the IAU recommended values for R$_0$ and $\Theta_0$ need a substantial 
revision. In particular the combination of 8.5 kpc and 220 \kms\, can be ruled
out with high confidence.

The weighted average of the four direct measurements for R$_0$ of 8.4 $\pm$ 0.4
kpc \citep{GhezSalimWeinberg2008}, 8.33 $\pm$ 0.35 kpc \citep{GillessenEisenhauerTrippe2009},  7.9$^{+0.8}_{-0.7}$ kpc \citep{ReidMentenZheng2009b}, and 8.4 
$\pm$ 0.6 kpc \cite{ReidMentenZheng2009} gives best value of 
\begin{eqnarray}
\nonumber \mathrm{R}_0 &=& 8.3 \pm 0.23~\mathrm{kpc}. 
\end{eqnarray}
\noindent This translates then into a rotation speed of 
\begin{eqnarray}
\nonumber \Theta_0 &=& \Theta_0/{\mathrm R}_0\times {\mathrm R}_0 = 28.79~\mathrm{km~s}^{-1}~\mathrm{kpc}^{-1} \times 8.3~\mathrm{kpc}\\
\nonumber &=& 239 \pm 7~\mathrm{km~s}^{-1}.
\end{eqnarray}
These values are almost identical to values presented by \cite{McMillan2011} 
and marginally consistent with the rotation speed of 224 $\pm$ 13 \kms\, 
provided by \cite{KoposovRixHogg2010} from modelling the stellar stream GD-1. 
For the old
Solar motion of V$_\odot$ = 5.25 \kms, $\Theta_0$ would increase to 246 $\pm$ 
7 \kms. The uncertainties are formal uncertainties and are dominated 
by the uncertainty in R$_0$. The true uncertainties are probably a factor of 
$\sqrt 2$ larger, since the two stellar orbit distances may have similar 
systematic errors (they are independent measurements with different 
instruments, but of the same source). 

A higher value of $\Theta_0$ is also supported 
by the observation of a retrograde-rotating and metal-poor component in the 
stellar halo of the Milky Way by \cite{DeasonBelokurovEvans2010} when assuming 
a value of $\Theta_0$ = 220 \kms. A value of $\Theta_0$ $\sim$ 240 
\kms\, would explain this apparent retrograde rotation.

\section{The Bar and Spiral Structure Legacy Survey}
\begin{figure}[!h]
\centering
{\includegraphics[width=80mm]{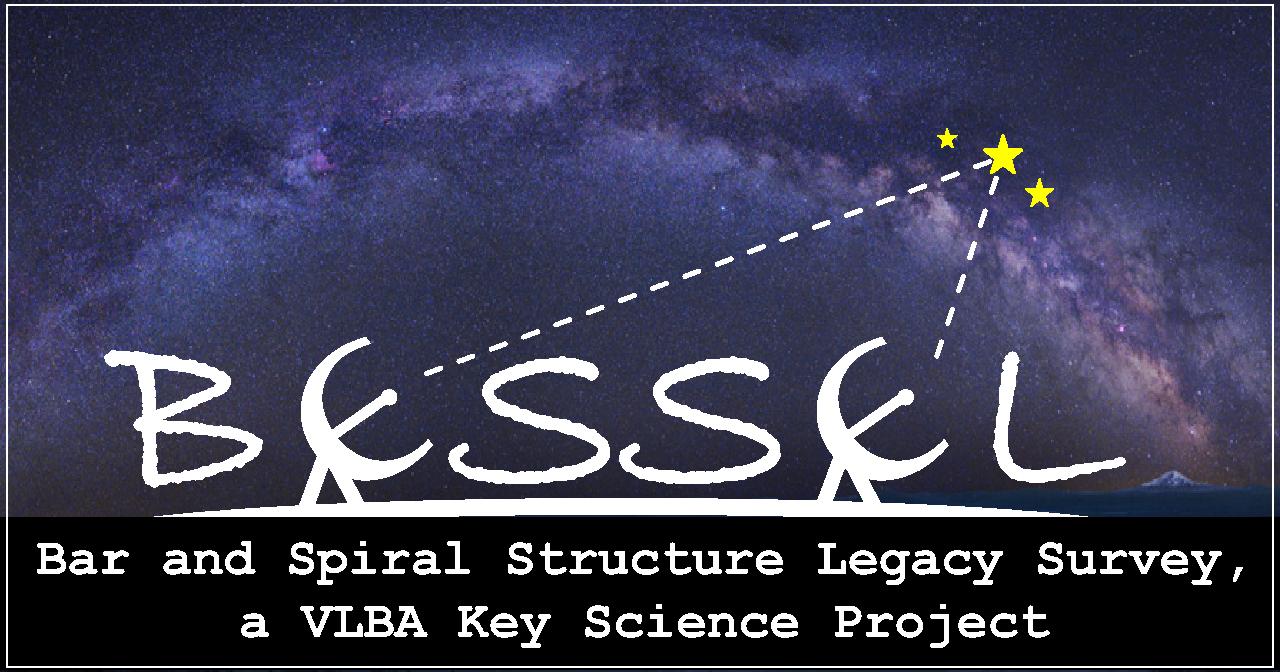}
\label{fig:current}}
\end{figure}
Motivated by these very encouraging results from only 18 sources, we have 
started a much larger project, the Bar and Spiral Structure Legacy (BeSSeL)\footnote{http://www.mpifr-bonn.mpg.de/staff/abrunthaler/BeSSeL/index.shtml} Survey,
a VLBA Key science project. The goal of BeSSeL, 
named in honor of Friedrich Willhelm Bessel who measured the first stellar 
parallax, is to measure accurate distances and proper motions of up to 400 high
mass star forming regions in the Milky Way between 2010 and 2015. This will 
result in a catalog of accurate distances to most Galactic high mass star 
forming regions visible from the northern hemisphere and very accurate 
measurements of fundamental parameters such as the distance to the Galactic 
center (R$_0$), the rotation velocity of the Milky Way ($\Theta_0$), and the 
rotation curve of the Milky Way. Additionally, maps of the maser distribution
from the first epoch of each source will be published on the BeSSeL website 
shortly after the observations.

\begin{figure*}
\centering
{\includegraphics[width=170mm]{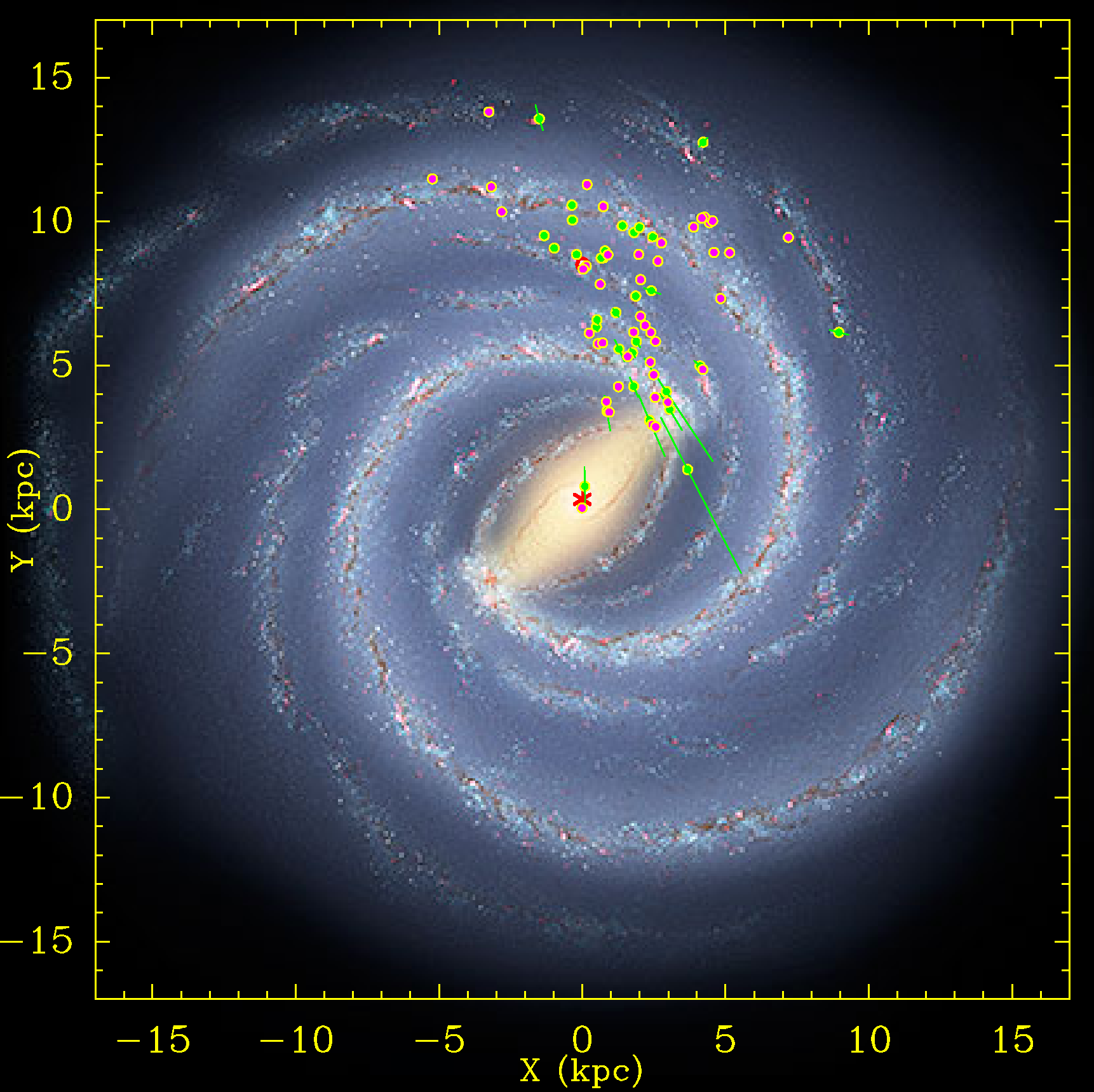}
\caption{Similar to Fig.~\ref{fig:sources}, but showing all sources currently measured (green), including unpublished sources, and all sources observed in the first year of BeSSeL (red), based on their kinematic distances.}
\label{fig:current}}
\end{figure*}

The BeSSeL Survey will first target 12.2 GHz methanol and 22 GHz water masers.
Once the VLBA is equipped with new receivers that also cover the 6.7 GHz 
methanol maser line (presumably in 2012) these masers will be also 
observed. In early 2010, preparatory surveys started with the Very Large Array 
to obtain accurate positions of the target water masers, and with the VLBA to
search for extragalactic background sources near the target maser sources 
\citep{ImmerBrunthalerReid2011}. The first parallax observations started in 
March 2010, and first results are expected in mid 2011 (see 
Fig.~\ref{fig:current}). In parallel, the VERA array will observe additional
H$_2$O and SiO masers throughout the Galaxy. Combined with complementary efforts
in the southern hemisphere with the Australian Long Baseline Array, this will
result into a detailed and accurate map of the spiral structure of the Milky 
Way. The superior sensitivity and the large field-of-view of the Square 
Kilometer Array, which will also cover the 6.7 GHz methanol maser line, will 
reach even higher astrometric accuracies for methanol masers, radio continuum
stars, and pulsars. This will result in a very detailed map of the the spiral 
stucture in the southern hemisphere. 

\acknowledgements
The Very Long Baseline Array (VLBA) is an instrument built and
operated by the National Radio Astronomy Observatory, a facility of
the National Science Foundation operated under cooperative agreement
by Associated Universities, Inc. The European VLBI Network is a joint facility 
of European, Chinese, South African and other radio astronomy institutes funded
by their national research councils.This work was partially funded by the 
ERC Advanced Investigator Grant GLOSTAR (247078). K.L.J.R. is funded by an 
ASI fellowship under contract number I/005/07/1.

\bibliography{aamnemonic,brunthal_refs}
\bibliographystyle{aa}

\end{document}